\documentclass[aps,pre,twocolumn,amsmath,amsfonts,amssymb,floatfix,superscriptaddress,nofootinbib]{revtex4}
\usepackage{mathrsfs} \usepackage{graphicx,color} \usepackage{bbm} \usepackage{multirow}
\usepackage{algorithm}

\usepackage{rotate}
\usepackage{epsfig}
\usepackage{psfrag}
\usepackage{mathtools}

\font\msytw=msbm9 scaled\magstep1

\let\a=\alpha  \let\g=\gamma \let\d=\delta
\let\e=\varepsilon \let\z=\zeta \let\h=\eta \let\k=\kappa
\let\l=\lambda     
\let\s=\sigma   
   \let\G=\Gamma
   
\let\Si=\Sigma   
  \let\io=\infty
\let\om=\omega

\def\DD{{\cal D}}

\def\de{\mathrm d}
\def\Di{D_{\rm int}}

\def\to{\rightarrow}

\def\ZZZ{\hbox{\msytw Z}}

\newcommand{\beq}{\begin{equation}} \newcommand{\eeq}{\end{equation}}
\newcommand{\wh}{\widehat}

\newcommand\be{\begin{equation}}
\newcommand\bea{\begin{eqnarray} \nonumber }
\newcommand\ee{\end{equation}}
\newcommand\eea{\end{eqnarray}}

\def\tpl{\tau_{\rm pl}}
\def\tel{\tau_{\rm fl}}

\begin{document}

\title{Spontaneous instabilities and stick-slip motion \\ in a generalized H\'ebraud-Lequeux model}

\author{Jean-Philippe Bouchaud}
\affiliation{CFM, 23 rue de l'Universit\'e, 75007 Paris, France, and Ecole Polytechnique, 91120 Palaiseau, France.}

\author{Stanislao Gualdi}
\affiliation{
Laboratoire de Math\'ematiques Appliqu\'ees aux Syst\`emes, CentraleSup\'elec, 92290 Ch\^atenay-Malabry, France
}

\author{Marco Tarzia}
\affiliation{
Universit\'e Pierre et Marie Curie - Paris 6, Laboratoire de Physique Th\'eorique de la Mati\`ere 
Condens\'ee, 4, Place Jussieu, 
Tour 12, 75252 Paris Cedex 05, France }

\author{Francesco Zamponi}
\affiliation{Laboratoire de Physique Th\'eorique,
\'Ecole Normale Sup\'erieure, UMR 8549 CNRS, 24 Rue Lhomond, 75231 Paris Cedex 05, France}

\begin{abstract}
We revisit the H\'ebraud-Lequeux (HL) model for the rheology of jammed materials and argue that a possibly important
time scale is missing from HL's initial specification. We show that our generalization of the HL model undergoes interesting oscillating instabilities 
for a wide range of parameters, which lead to intermittent, stick-slip flows under constant shear rate. The instability 
we find is akin to the synchronization transition of coupled elements that arises in many different contexts (neurons, fireflies, financial bankruptcies, etc.).
We hope that our scenario could shed light on the commonly observed intermittent, serrated flows of glassy materials under shear.
\end{abstract}

\maketitle


\section{Introduction}

Jammed materials are like Tolstoy's happy families: they all behave much in the same way. The (non-linear) rheological properties of very different 
materials such as metallic glasses, foams, dense emulsions or granular packings are remarkably similar. Typically, these materials 
creep very slowly below
a certain yield stress, and exhibit interesting dynamical features such as ``avalanches''~\cite{baret,OH14} (perhaps similar, at a different scale, to real earthquakes~\cite{chen}) and
intermittent, stick-slip like motion, which are still poorly understood -- as is the transition from ``liquid'' flows to ``jammed'' flows~\cite{LDW12}. 
At the microscopic
scale all of these materials have very different
properties and can barely be described by one single
model. However on a more coarse grained level, when for
example considering rheological or plasticity features, these
materials show very similar properties.
Such universality has motivated researchers to propose simple phenomenological models to account for them. 
It is now well accepted 
that the elementary physical process at play is the yielding of small regions of the material~\cite{STZ1,STZ2,STZa,STZb,STZ3,STZ4,STZ5,STZ6,STZ7} 
(the so-called shear transformation zones, STZ), when the local shear stress exceeds 
some (possibly random) threshold. 
These STZ are localized but display long-range elastic interactions~\cite{picard} that can trigger plastic instabilities.
Among the most popular approaches to describe the stochastic collective evolution of these STZ is the H\'ebraud-Lequeux (HL) model 
\cite{HL98}, which is a very simple, mean-field description of the elastic interaction between such STZ, in particular how the instability of one of them can trigger 
the instability of many others (see also Refs.~\cite{SLHC97,So06,JLR14,LELC15,coussot} for alternative descriptions and 
\cite{review} for a recent review on this topic). 
The HL model, in spite of its simplicity, leads to highly 
non-trivial results, in particular a phase transition between a jammed, arrested phase and a liquid, flowing phase, with a non-linear Herschel-Bulkley law~\cite{HB} 
for small shear rate in the jammed phase. This model has attracted a renewed interest in the recent years, focusing, inter-alia, on a detailed comparison with 
the competing Soft-Glassy-Rheology (SGR) models \cite{ABMB15}, on avalanche-like dynamics \cite{Ja15}, and on a generalization of HL 
that appropriately accounts for the power-law decay 
of elastic interactions between STZ~\cite{LW15,LC07}, which deeply modifies the mathematical structure of the exact solution of the model. 

In this paper, we revisit the original HL model and argue that some important physical time-scales are missing from its  original formulation. 
Although many of the predictions of HL are unaffected by these modifications, we find that an oscillating instability can appear in some regions 
of parameter space. We find in particular that the ``liquid'' phase at zero shear rate is unstable and becomes, close enough the the 
jammed phase, spontaneously oscillating between a self-sustained liquid and a glass.
This instability persists when the shear rate is weak enough, and 
leads, in physical terms, to an intermittent stick-slip flow (possibly related to so-called serrated flow). 
Interestingly, oscillatory instabilities also arise in some variants of the SGR models~\cite{cates1}, or within simple phenomenological 
constitutive equations for shear-thickening materials~\cite{cates4}. However, the underlying physical mechanisms leading to instabilities 
are quite different from the one discussed in the present paper.

The intuition behind our instability in fact comes from a very similar model that we recently studied in the context of economic crisis ``waves'' and collective 
bank defaults \cite{GTZB15,GBCTZ15}. In all these cases, the common phenomenon is that the instability/default/bankruptcy of a single element can trigger the 
instabilities of others through interactions. The somewhat surprising feature, however, in that such a coupling can lead to the synchronization of randomly evolving entities, and genuine oscillations -- see 
\cite{St03,GBCTZ15} and references therein.

The outline of the paper is as follows: we first generalize the HL model such as to describe more accurately what happens when a single entity fails. We then
show that the zero shear stationary state computed by HL is linearly unstable to some oscillatory mode in a region of parameters, in particular close to the
jamming transition. We then extend our calculation to non-zero shear rates and find the ``phase diagram'' of the model, in particular we highlight the region where stick-slip motion is 
expected. We finally conclude and discuss possible extensions of our calculations.

\section{The model}

Let us recall the basic tenets of the HL model~\cite{HL98} and introduce the physical items that
we believe are missing from the original framework. Following HL, the material is divided in $N$ regions (or ``elements'') labeled by an index
$i = 1,\cdots,N$, each characterized by a local stress $\s_i$.
The stress $\s_i$ of each element performs a random walk with:
\begin{itemize}
\item a drift $G_0 \dot\g$ due to the external shear $\dot \g$ that loads elastic energy ($G_0$ is the elastic shear modulus),
\item a time-dependent diffusion $D_t$ due to intrinsic noise and elastic interactions between elements.
\end{itemize}
Whenever $|\s_i | \geq \s_c$, the $i$-th element may become unstable, and becomes so at a ``plastic'' rate $1/\tpl$.
As soon as the element becomes unstable (it is then in a ``fluid'' state), HL postulates that it re-jams (with $\s=0$)
immediately in a state of zero stress. We rather assume that it remains in its fluid
state during a typical time scale $\tel$; in such a state, the element cannot contribute to the shear stress
(see also Fig. 1 in Ref.~\cite{MBB12}, where $\tel$ has been called $\tau_{\rm el}$). 
With rate $1/\tel$, the element re-jams with $\s=0$, as in  HL.\footnote{This could actually be generalized to the case 
where the stress drops to a finite fraction of the initial stress (as in \cite{Puosi}), or to a non-trivial distribution
of initial stresses; the instability reported below survives provided the width of that distribution remains small.}  
This is a very important ingredient for a correct 
physical interpretation of the model~\cite{MBB12,JLR14}: note that once an element becomes unstable, 
its dynamics cannot be described by the same drift-diffusion mechanism. 
On the contrary, when an element becomes unstable, it contributes to the total plastic activity at time $t$, called $\G_t$,
and increases the diffusion of all other elements identically, in a mean-field way, exactly as in the HL model (see \cite{LW15} for an improved description). However, we will assume that
the plastic activity at time $t$ affects the diffusion coefficient $D_t$ with some delay, instead of instantaneously as HL assumed.
This reflects, at the mean-field level, the finite propagation speed of information through the sample.

\subsection{Mathematical definition of the model}

Our modified model can be described by the same Fokker-Planck setting as used by HL~\cite{HL98}.
The evolution of the (unnormalized\footnote{The reason why $P(\s,t)$ is not normalised is that 
it does not include the
fraction $(1-\phi_t)$ of elements that are in the fluid state, see Eq.~\eqref{eq:2}.}) 
probability $P(\s,t)$ to find an element with stress $\s$ at time $t$ is given by:
\be\label{eq:FP}
\begin{split}
\dot{P}(\s,t) &= D_t P''(\s,t) - G_0 \dot\g P'(\s,t) \\ &+ J_t\delta(\s) - \frac1\tpl P(\s,t) H(|\s| - \s_c) \ ,
\end{split}\ee
where $H(x)$ is the Heaviside step function, and
with:
\beq\label{eq:2}
\begin{split}
J_t &= \frac1\tel (1-\phi_t) \ , \\
\phi_t &= \int_{-\io}^{\infty} d\s \, P(\s,t) \ , \\
\G_t &= \frac1\tpl \int_{|\s|\geq \s_c} d\s \, |\s| P(\s,t) \ , \\
D_t &= \Di + \a \omega \int_{-\io}^t ds \,\, e^{-\omega (t-s)} \G_s  \ .
\end{split}
\eeq
Here, $\phi_t$ is the fraction of jammed elements that contribute to the elastic stress and $J_t$ is 
the flux of elements that re-jam between
$t$ and $t + {\rm d}t$, initially at zero stress. 
One can assume for simplicity\footnote{
As it will be clear in the following, the stress $\Si_t$ is an observable derived from the model but it does not
enter into the main equations. Therefore, its precise form does not affect the main conclusions of this paper
about the phase diagram of the model. One could thus choose a different form (e.g. a viscoelastic one)
for the stress contribution of the fluidized elements.
} 
that the fraction of fluidized elements, $(1 - \phi_t)$, contributes a viscous stress proportional to $\dot\g$.
Hence, the total stress is
\beq\label{eq:stress}
\Si_t = \int_{-\io}^{\infty} d\s \, \s \, P(\s,t) + (1 - \phi_t)\h \dot\g \ ,
\eeq
where $\h$ is the microscopic viscosity of the fluid elements.
Note that
$\phi_t$ satisfies the equation:
\beq\label{eq:dotphi}
\dot \phi_t = \frac1\tel (1-\phi_t) - \frac1\tpl \int_{|\s|\geq \s_c} d\s P(\s,t) \ .
\eeq
The equation for $D_t$ means that yielding events, when they happen, impose an extra random stress onto other elements but with 
some time lag that we model as an exponential kernel, with a coupling constant $\a$. $\Di$ is an intrinsic noise term (for example temperature) that
we will choose to be very small in the following. Furthermore we choose $G_0=1$ without loss of generality\footnote{
Note that $G_0$ only appears in Eq.~\eqref{eq:FP} together with $\dot\gamma$.
Hence, choosing $G_0=1$ is equivalent to a change of notation, $G_0 \dot\gamma \to \dot\gamma$; it is not a choice of units.
}.
The control parameters for this model are $\s_c, \dot\g, \tel, \tpl, \Di, \a, \omega$.
Because one can fix arbitrarily the units of $\s$ and of $t$, there are in fact four plus one independent adimensional control parameters:
for example $\wh \g := \dot\g \tel/\s_c$, $\a$, $\tpl/\tel$, $\widehat\om := \omega\tel$, and finally $\text{Pe}:=\dot\g\s_c/\Di \to \infty$ throughout this paper. 

\subsection{Connection with the H\'ebraud-Lequeux model}

The standard HL model is recovered if we set $\Di=0$, $\tel=0$ and $\omega=\io$, i.e: no intrinsic noise, fluid regions re-jam instantaneously, 
and yielding elements affect the stress of other regions instantaneously as well. In the HL limit $\tel=0$, 
we therefore have $\dot \phi_t=0$ and $\phi_t \equiv 1$,
and from Eq.~\eqref{eq:dotphi} we have
$J_t = \frac1\tpl \int_{|\s|\geq \s_c} d\s P(\s,t)$. For $\Di=0$ and $\omega\to\io$, we indeed recover HL's prescription $D_t = \a \G_t$.
Summarizing, we obtain Eq.~\eqref{eq:FP} with
\beq
\begin{split}
J_t &=\frac1\tpl \int_{|\s|\geq \s_c} d\s P(\s,t) \ ,\\
D_t &= \a \G_t =  \frac\a\tpl \int_{|\s|\geq \s_c} d\s |\s| P(\s,t) \ . 
\end{split}\eeq
This coincides with the HL model, with the only difference that $\G_t$ is defined without the $|\s|$. 
Although this does not make any important difference in the properties of the model, in particular the instability described below, 
we argue that our definition has a more precise physical interpretation, 
since the plastic activity should be the average of $|\s|$.
In conclusion, the presence of the additional time scales $\tel$ and $1/\omega$ is the main difference between our model and HL,
and we will see that they play a crucial role in the physics of the model.

\subsection{The limit $\tpl \to 0$}

For the ease of analytical calculations, we will set in the following $\tpl = 0$, which means that an element becomes unstable immediately 
when $|\s| > \s_c$, while keeping a finite fluid lifetime $\tel$. 
The HL model corresponds to the opposite limit $\tpl/\tel \to \io$, but
it is reasonable to expect that in many applications
the ratio $\tpl/\tel$ can take very different values, from quite small to quite large.
In this paper we examine analytically the limit $\tpl/\tel \ll 1$, and numerically the regime where $\tpl/\tel \sim 1$.
For small $\tpl/\tel$ we find
an oscillating instability, which disappears when $\tpl/\tel$ exceeds a (parameter dependent) threshold value, simply because the 
synchronisation effect reported below cannot set in. So strictly speaking, the
instability reported here is absent in the original HL setting.

When $\tpl = 0$, Eq.~\eqref{eq:FP} is simply complemented by an absorbing boundary condition at $|\s| = \s_c$, 
implying $P(|\s|=\s_c)=0$.
Furthermore, the plastic activity is given by the number of regions that become unstable per unit time, hence in Eq.~\eqref{eq:2} $\G_t$
is given
by the outgoing flux at $\pm \s_c$, i.e.:
\beq\label{eq:Gtpl0}
\G_t = \s_c D_t [ P'(-\s_c,t) - P'(\s_c,t) ] \ .
\eeq
Note that in this limit the difference in our definition of $\G_t$ with respect to HL
becomes irrelevant, because $|\s|=\s_c$ is the only contribution to $\G_t$.
For simplicity, we also set $\s_c=1$ in the following. 
 
\section{Vanishing shear rate}
\label{sec:gammazero}

We will consider first the situation with no external shear, i.e. $\dot\g=0$. 
It turns out that the interesting limit corresponds to the case when the intrinsic activity tends to zero, i.e. $\Di \to 0$, as in the HL model. The only two adimensional control parameters left are thus $\a$ and $\widehat\om := \omega \tel$.

Note that, as we will show in the following (Sect.~\ref{sec:stashear}), for $\a < 1/2$ and $\Di=0$
there is a history-dependent jammed phase in which the material can self-sustain, 
in the $\dot\gamma = 0$ limit, any finite stress within the interval $[-\sigma_c,+\sigma_c]$. In this case, setting directly
$\dot\gamma = 0$ as we do below corresponds to a particular choice of history such
that stationary stress distribution is symmetric and the average stress is zero.
The main point of this section is to illustrate our calculation in the simplest possible case.

\subsection{The stationary solution: a jamming transition}

\begin{figure}[t]
\includegraphics[width=.9\columnwidth]{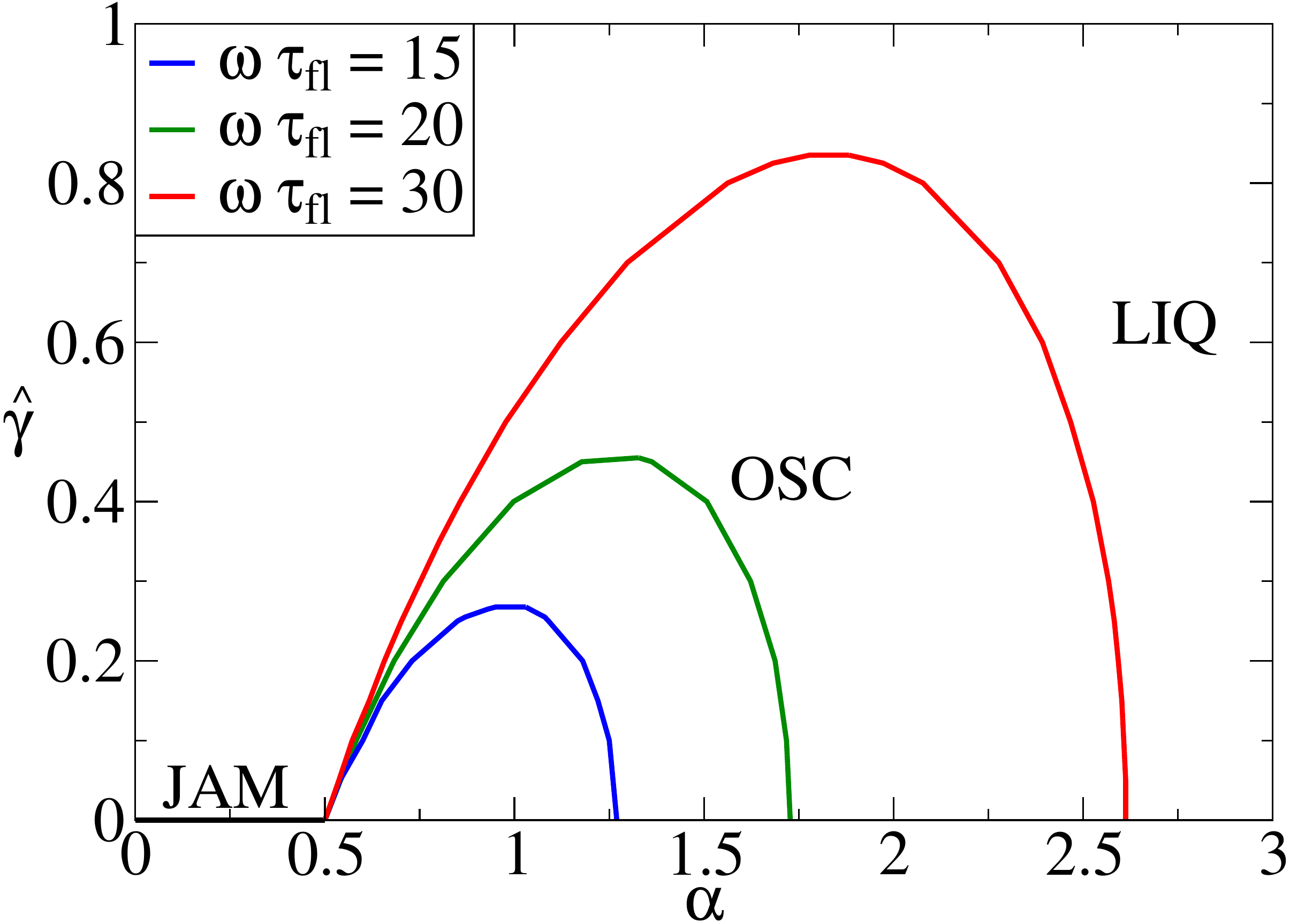}
\includegraphics[width=.9\columnwidth]{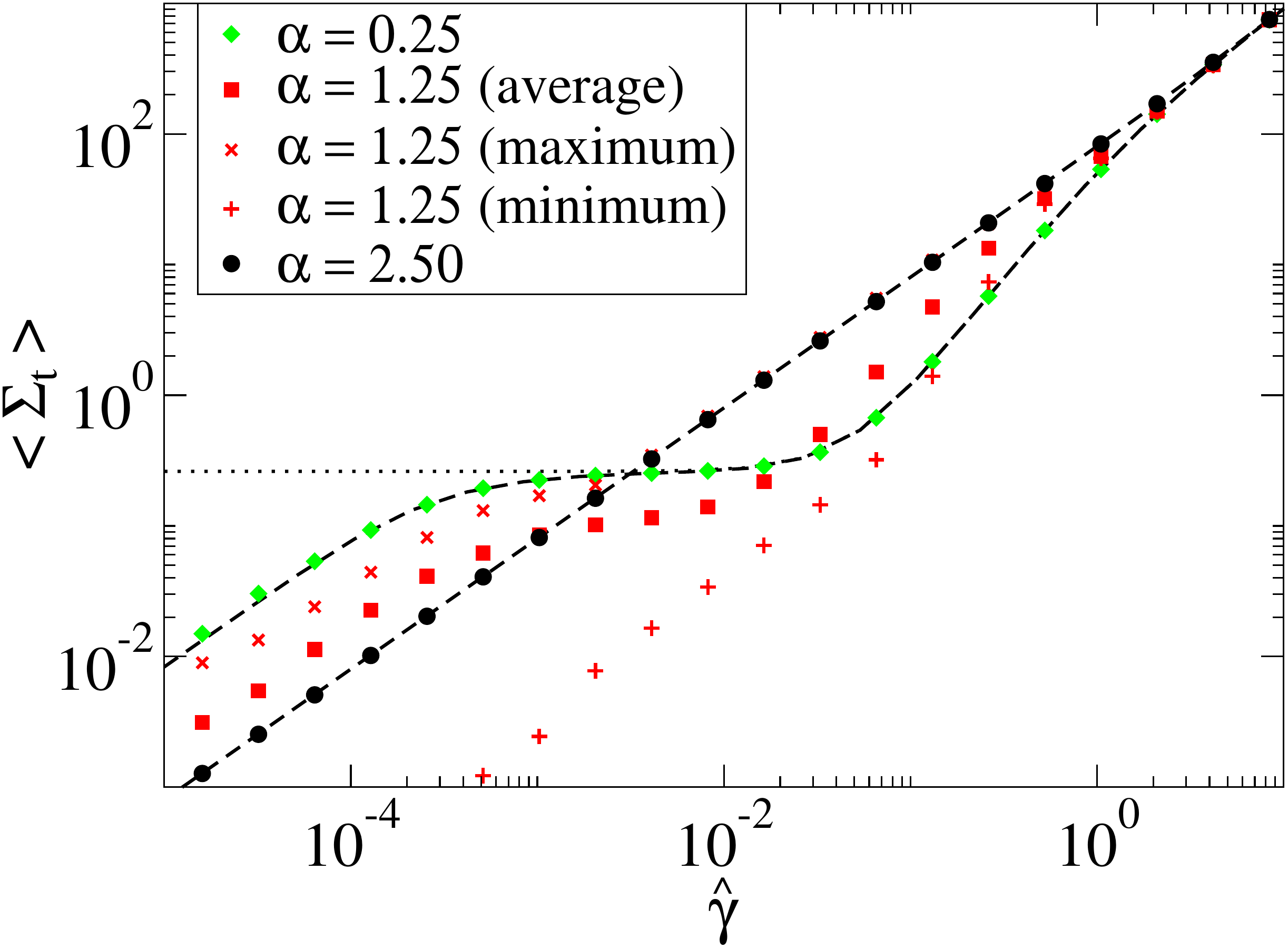}
\caption{
\emph{Top}: 
Phase diagram of the model in the $(\a,\wh\g )$ plane (recall that $\wh\g = \dot\g \tel$) obtained from the solutions of Eq.~\eqref{nonzerosol}
with $D_i\to 0$ and different values of $\widehat{\omega}=\omega\tel$.
In the liquid (LIQ) phase, the system is stationary, and the flow curve is basically Newtonian in the whole range of $\wh\g$. In the
jammed (JAM) phase, the system has a finite yield stress for $\wh\g\to0$. Finally, in the oscillating (OSC) phase, the stress is an oscillating
function of time. 
\emph{Bottom}:
Flow curves displaying the average stress versus strain for $\wh\omega=20$, $\Di = 0.005$, and 
three different values of $\a$ corresponding to the jammed, oscillating, and liquid
phases. Points correspond to numerical results while lines correspond to analytical results for the stationary state.
For $\a=2.5$ the system is liquid and the flow curve is Newtonian in all the regimes of $\wh\g$. 
For $\a=0.25$ the system is jammed for $\wh\g\to0$ and the stress tends to a finite value (dotted line), 
but in the simulation due to finite $\Di$ we also see a Newtonian regime at small $\wh\g$ (points and dashed line).
For $\a=1.25$ the system is oscillating for small $\wh\g$: in this case we report the minimum, average and
maximum value of the stress (see Fig.~\ref{fig:2} for an example).
}
\label{fig:1}
\end{figure}

We thus look for a stationary solution of Eq.~\eqref{eq:FP} with $P(\s,t)=P_0(\s)$, $J_t = J_0$, etc.,
which in this case is particularly simple and reads:
\be\label{eq:Pnonshear}
P_0(\s > 0) = A ( 1 -\s ) \ , \qquad P_0(-\s)=P_0(\s) \ ,
\ee
with the discontinuity of the slope at the origin related to $J_0$ by Eq.~\eqref{eq:FP} as $2 D_0 A = J_0$.
The fraction of active elements $\phi_0 = \int_{-1}^1 d\s P_0(\s)$ is simply given by $A$, so that from Eq.~\eqref{eq:2}
\be
J_0 = \frac1{\tel} (1 - A) \hskip10pt  \longrightarrow \hskip10pt A = \frac{1}{2D_0\tel + 1} \ .
\ee
Eq.~\eqref{eq:Gtpl0} gives $\G_0 = 2 A D_0$ and from Eq.~\eqref{eq:FP} we get $D_0 = \Di + \a \G_0
= \Di + 2 \a A D_0$. We thus obtain
a self-consistent equation for the stationary value of the
diffusion coefficient
\be
2 D_0^2 \tel - 2 D_0 (\d\a + \Di \tel ) - \Di=0 \ ,
\ee
where $\d\a = \a - 1/2$.
This second degree equation has two possible solutions that read, in the limit $\Di \to 0$:
\be\label{eq:D0_g0}
D_0^- = -\frac{\Di}{ 2\d \a} \ ; \qquad D_0^+ = \frac{2\d\a }{2\tel}  + \frac{\a \Di \tel}{4\d\a} \ ,
\ee
where clearly the first solution holds for $\a <1/2$ and the second holds for $\a > 1/2$. When $\Di \to 0$ there is, as first noted by HL,
a transition between a fluid phase $\a > 1/2$ where activity is self sustained $D_0^+ > 0$, and a jammed phase $\a < 1/2$ where 
activity is absent, $D_0^- = 0$, see Fig.~\ref{fig:1} and Ref.~\cite{ABMB15}. 
In the jammed phase, one has $A=1$ and in the liquid phase, $A=1/(2 \a)$, to leading order in~$\Di$.

\subsection{Linear stability analysis}

Following~\cite{GBCTZ15}, we are interested in studying the stability of the stationary solution with respect to a linear perturbation.
We thus write $P(\s,t) = P_0(\s) + \e P_1(\s,t)$, $D_t = D_0 + \e D_1(t)$, $\phi_t = \phi_0 + \e \phi_1(t)$.
Linearizing Eq.~\eqref{eq:FP} we get
\be\label{HL2}
\dot{P}_1(\s,t) = D_0 P_1''(\s,t) + D_1(t) P_0''(\s) - \frac1{\tel} \phi_1(t) \delta(\s) \ ,
\ee
with the boundary conditions $P_1(\s=\pm1,t)  = 0$.
Therefore,
\be\label{eq:P1}
\begin{split}
P_1(\s,t) &=\int_0^\infty d\tau \int_{-1}^1 d\bar\s \, G(\s,t| \bar\s ,t-\tau) \\ 
&\times \left[D_1(t-\tau) P_0''(\bar\s) - \frac1{\tel} \phi_1(t-\tau) \delta(\bar\s)\right] \ .
\end{split}\ee
where $P_0''(\s) = - 2A \delta(\s)$, and $G$ is the random walk propagator in the interval $[-1,+1]$.

Combining Eqs.~\eqref{eq:2} and \eqref{eq:Gtpl0}, setting $\Di=0$ in the linear regime we obtain
an equation for the diffusion constant $D_1(t)$:
\be\label{eq:LD1}
\begin{split}
D_1(t) &=  \a \om \int_{-\io}^t ds D_0 \left[P_1'(-1,s) - P_1'(1,s)\right] e^{-\omega (t-s)} \\
& + 2 A \a  \om \int_{-\io}^t ds D_1(s) e^{-\omega (t-s)} \ .
\end{split}\ee
Exponentials of time, 
with $D_1(t) = \DD e^{\l t}$, $\phi_1(t) = \Phi  e^{\l t}$, $P_1'(-1,t)=-P_1'(1,t)=\Pi e^{\l t}$, 
are consistent solutions of the linearized equations.
We 
find from Eq.~\eqref{eq:LD1}:
\be\label{eq:C1}
\DD = \frac{ 2 \a \om D_0}{\om (1- 2 \a A) + \l} \Pi \ .
\ee

In order to compute $P_1(\s,t)$ simply, 
one can notice that $H_{\bar\s}(\s,\l) = \int_0^\infty d\tau e^{-\l \tau} G(\s,\tau|\bar\s,0)$ obeys the following 
partial differential equation:
\be
D_0 H_{\bar\s}''(\s,\l) - \l H_{\bar\s}(\s, \l) = - \delta(\s-{\bar\s}) \ ,
\ee
with boundary condition $H_{\bar\s}(\s = \pm 1, \l )= 0$. Specializing to ${\bar\s} = 0$, one finds:
\be\begin{split}
& H(\s > 0,\l) = \widehat A \left[ e^{-\k \s} - e^{-2\k} e^{\k \s} \right] \ , \\
& H(-\s, \l) = H(\s, \l)
\end{split}\ee
with $\k^2 = \l/D_0$. The discontinuity of the first derivative at the origin gives:
\be
2 D_0 \widehat A  = \frac{1}{\k(1 + e^{-2\k})} \ ,
\ee
so finally $H'(-1,\l) = - H'(1,\l) = 1/(2D_0 \cosh \k)$.
Plugging this in the general expression for $P_1(\s,t)$ in Eq.~\eqref{eq:P1}, 
taking the derivative with respect to $\s$ and setting $\s=-1$ finally leads to:
\be\label{eq:C2}
\Pi = - \frac{1}{2D_0 \cosh \k} \left[2A \DD + \frac1{\tel} \Phi \right].
\ee
Finally, integrating Eq. (\ref{HL2}) over $\s$ leads to:
\be\label{eq:C3}
\l \Phi = - \frac1{\tel} \Phi - 2D_0 \Pi - 2 \DD A \ .
\ee
In the liquid phase ($D_0 > 0$ and $A= 1/(2 \a)$)
the solvability condition of the system of Eqs.~\eqref{eq:C1}, \eqref{eq:C2}, \eqref{eq:C3}
 relating $\DD, \Pi$ and $\Phi$ finally reads, in terms of $\widehat \l = \l \tel$ and 
$\widehat \om = \om \tel$:
\be
1 + \widehat \l = \frac{\widehat \om + \widehat \l}{\widehat \om + \widehat \l \cosh \sqrt{\frac{\widehat \l}{\d\a }}} \ ,
\ee
or equivalently (for $\widehat\l \neq 0$)
\beq\label{eq:stab1}
\widehat\om = 1 - (1+\widehat\l) \cosh \sqrt{\frac{\widehat \l}{\d\a}} \ .
\eeq
If all the solutions of Eq.~\eqref{eq:stab1} have a negative real part, the stationary solution
is stable, while it is unstable otherwise. 
This condition defines the transition point $\a_{\rm c}(\widehat\om)$.
Note that for $\a\to\io$ we obtain from Eq.~\eqref{eq:stab1} that $\widehat\l = -\widehat\om$ and therefore
the stationary state is stable.

\subsection{Large $\om$ analysis}

Eq.~\eqref{eq:stab1} can be analyzed in the limit of large $\om$ (that corresponds to the HL case where stress propagation is
instantaneous throughout the sample), assuming that $\a_{\rm c}(\widehat\om)$
and $\widehat\l$ are also large (we will see that they are proportional to $\wh\om$).
We obtain
\beq\label{eq:stab2}
\widehat\om = -\widehat\l \cosh \sqrt{\frac{\widehat \l}{\a}} \ .
\eeq
Let us assume that at the transition there is a single unstable mode whose real part crosses zero continuously.
Thus $\widehat \l = i \l_{\rm I}$ is pure imaginary at $\a_{\rm c}$.
In this case, for the r.h.s. of Eq.~\eqref{eq:stab2} to be real we need $\cos \sqrt{\l_{\rm I}/(2\a)} =0$
which implies $\l_{\rm I} = \a \pi^2 (1+2n)^2/2$ with $n \in \ZZZ$. Then we get
\beq\label{eq:stab3}
\frac{\widehat\om}\a = (-1)^n \frac{\pi^2}2 (1+2n)^2 \sinh \left( \frac{\pi}{2} (1+2n)     \right) \ .
\eeq
Because the system is stable at $\a\to\io$, 
the stability boundary corresponds to the largest value of $\a$ for which a mode is unstable,
and thus to the smallest value of the r.h.s. of Eq.~\eqref{eq:stab3}, which is assumed for $n=0$
and gives
\beq
\frac{\a_{\rm c}(\widehat\om)}{\widehat\om} = \frac{2}{\pi^2 \sinh \left( \frac{\pi}{2} \right)} =0.088055\ldots \ ,
\eeq
in excellent agreement with the numerical results of Fig.~\ref{fig:1} when $\widehat\om$ is large enough (for example, 
$\widehat\om = 30$ should correspond to $\a_{\rm c} \approx 2.64$).

\subsection{Summary}

Let us summarize the content of this section. We discussed the case of vanishing shear rate, $\dot\g=0$, and intrinsic activity, $\Di \to 0$,
assuming $\tpl \ll \tel$. The control parameters are $\a$ and $\widehat\om = \om\tel$. 
We found that:
\begin{itemize}
\item For $\a<1/2$, the system is jammed: no motion occurs as $D_t \equiv 0$.
\item For $\a > \a_{\rm c}(\widehat\om)\approx \widehat\om \times 0.088055\ldots $, there is a stationary ``liquid'' phase with self-sustained activity, $D_t \equiv D_0 > 0$.
\item For $1/2 < \a < \a_{\rm c}(\widehat\om)$ the liquid phase is in fact unstable; in this case, as shown in Fig.~\ref{fig:2}, the system undergoes spontaneous 
oscillations between periods where the number of fluid elements is large and periods where it is small. When an external shear is applied, this state will 
correspond to intermittent, stick-slip motion.
\end{itemize}
Let us emphasize that the liquid phase of the model 
is unphysical in the strict limit $\Di=0$ and $\dot\g = 0$, because no motion 
is possible in this case. However, one can think that $\Di$ 
is very small but finite, modeling some external noise (e.g. a small temperature), or that $\dot\g\to 0^+$. 
As we discuss next, the more interesting case $\dot\g>0$ is indeed perfectly regular in the limit 
where $\dot\g\to 0$.

\section{Non-zero shear rate}

We now generalize the calculation to a non-zero external shear, $\dot \gamma \neq 0$, still in the limit $\Di \to 0$.
In addition to $\a$ and $\widehat\om = \omega \tel$, there is now a third control parameter $\widehat\g = \dot\g \tel$.
We also define $\zeta= \dot\g \sigma_c/D_0$, 
which is an adimensional ``renormalized P\'eclet number'' 
quantifying the importance of drift with respect to the steady-state diffusion constant in Eq.~\eqref{eq:FP}. Remember however that we set $\sigma_c=1$
henceforth.

\subsection{The stationary case}
\label{sec:stashear}

The stationary solution now reads:
\be\label{eq:Pshear}
\begin{split}
P_0(\s > 0) &= A ( 1 -  e^{-\z (1-\s)}) \ , \\
P_0(\s < 0) &= -A e^{-\zeta} ( 1 -  e^{\z (1+\s)}) \ .
\end{split}\ee
Integrating Eq.~\eqref{eq:FP} one can relate the discontinuity of the slope at the origin to $J_0$ as
$A =  \frac{J_0}{\dot\g} \frac{1}{1 + e^{-\z}}$. Moreover,
the fraction of active elements is $\phi_0=A(1 - e^{-\z})$, and
$J_0 =  (1 - \phi_0) /\tel$. We thus have three equations for $J_0$, $\phi_0$
and $A$ and the solution is
\be
J_0 =  \frac{ \dot\g}{\dot\g \tel + \tanh(\z/2)} \ ,
\ee
from which we obtain $\phi_0$ and $A$ thus completing the calculation of $P_0(\s)$.

From $P_0(\s)$ we obtain the self-consistent equation obeyed by $D_0 = \dot\g/\z$.
Combining Eqs.~\eqref{eq:2} and \eqref{eq:Gtpl0}, we have
$D_0 = \Di + \a \G_0 = \Di + \a D_0 [P_0'(-1) - P_0'(1) ]$.
For $\Di\to0$ we obtain:
\be
\frac{D_0}{\dot\g} =
\frac1\z = \frac{\a }{\widehat\g + \tanh(\z/2)} \ ,
\ee
which is more conveniently written as
\beq\label{eq:zetaeq}
\widehat\g + \tanh(\z/2) = \a \z \ .
\eeq
One is now always in a ``liquid'' phase with a finite diffusion constant $D_0 > 0$, induced by the shear rate. 

We can also compute the average stress. We obtain from Eq.~\eqref{eq:stress}
\beq
\Si_0 = \frac{\z - 2 + (\z+2) e^{-\z}}{2 \a \z^2 (1 + e^{-\z})} + \widehat\h \frac{\widehat\g^2}{\a \z} \ ,
\eeq
where $\widehat\h = \h/\tel$.
Note that for large $\wh\g$, Eq.~\eqref{eq:zetaeq} gives $\z \sim \wh\g/\a$ and thus $\Si_0 \sim \wh\h \, \wh\g$ is
always Newtonian, begin dominated by the fluidized elements.
A plot of $\Si_0$ versus $\wh\g$ for several values of $\a$ is reported in Fig.~\ref{fig:1}.

Note that the behavior at small $\dot\g$ is  
very different when $\a > 1/2$ or when $\a < 1/2$. In the limit $\wh\g \to 0$:
\begin{itemize}
\item
For $\a > 1/2$, it is easy to see graphically that Eq.~\eqref{eq:zetaeq} has a unique solution for $\z$ for all $\widehat\g$,
and that $\z \sim \wh\g/\d\a \to 0$. Hence, Eq.~\eqref{eq:Pshear} reduces to Eq.~\eqref{eq:Pnonshear}. Furthermore,
\beq
D_0 = \wh\g/(\z \tel)\to  \d\a/\tel =D_0^+
\eeq
as in Eq.~\eqref{eq:D0_g0},
while the stress
\beq
\Si_0 \sim \frac{\wh\g}{\a} \left( \frac1{24\d\a} + \wh\h \d\a \right) 
\eeq
is Newtonian, but with a very different prefactor from the large $\wh\g$ regime.
\item
For $\a \leq 1/2$, one finds instead that Eq.~\eqref{eq:zetaeq} has multiple solutions when $\wh\g$ is small enough. Yet at large
$\wh\g$ the solution is unique and reducing $\wh\g$ gradually one sees that  $\forall \wh\g > 0$ the largest solution for $\z$ has to be chosen.
This solution for $\z$ remains finite for $\dot\g\to0$. Consequently, Eq.~\eqref{eq:Pshear} does not reduce to Eq.~\eqref{eq:Pnonshear}, which
shows that for $\a \leq 1/2$ one cannot set $\dot\g=0$ directly because the limit is singular. 
Setting directly $\dot\gamma  = 0$ as in Sec.~\ref{sec:gammazero} corresponds
to choosing a particular solution, corresponding to $\Sigma_0=0$, 
which is therefore not the physically natural solution, 
obtained by preparing the system under shear and reducing progressively $\dot\gamma$ to zero.
In the latter case, $D_0 \propto \wh\g$ and $\Si_0$ remains finite when $\dot\g \to 0$: the system is jammed, has no plastic activity and can sustain a
finite stress even for $\dot\g=0$.
\end{itemize}
Note that $\z\to 0$ for $\a \to 1/2^-$. Thus $D_0 = D_0^+ + O(\dot\g^2)$ for $\a>1/2$ -- see also~\cite{ABMB15}.

\subsection{Linear stability analysis}

We linearize once again Eq.~\eqref{eq:FP} with
$P(\s,t) = P_0(\s) + \e P_1(\s,t)$, $D_t = D_0 + \e D_1(t)$, $\phi_t = \phi_0 + \e \phi_1(t)$.
The linearized equation reads:
\be\label{HL3}\begin{split}
\dot{P}_1(\s,t) & = D_0 P_1''(\s,t) -\dot\g P_1'(\s,t) \\ &+ D_1(t) P_0''(\s) - \frac1{\tel} \phi_1(t) \delta(\s) \ ,
\end{split}\ee
with the boundary conditions $P_1(\s=\pm 1 ,t) = 0$.
Therefore 
\be\label{eq:P1shear}
\begin{split}
P_1(\s,t) &=\int_0^\infty d\tau \int_{-1}^1 d\bar\s \, G_{\dot\g}(\s,t| \bar\s ,t-\tau) \\ 
&\times \left[D_1(t-\tau) P_0''(\bar\s) - \frac1{\tel} \phi_1(t-\tau) \delta(\bar\s)\right] \ ,
\end{split}\ee
where $G_{\dot\g}$ is the biased random walk propagator in the interval $[-1,+1]$.
The diffusion constant $D_1(t)$ satisfies:
\be\label{eq:D1shear}
\begin{split}
D_1(t) &=  \a \om \int_{-\io}^t ds D_0 \left[P_1'(-1,s) - P_1'(1,s)\right] e^{-\omega (t-s)} \\
& + \frac{J_0}{D_0} \a  \om \int_{-\io}^t ds D_1(s) e^{-\omega (t-s)} \ .
\end{split}\ee
Assuming exponential (in time) solutions with $D_1(t) = \DD e^{\l t}$, $\phi_1(t) = \Phi  e^{\l t}$, $P_1'(\pm 1,t)=\mp \Pi_\pm e^{\l t}$, we 
find from Eq.~\eqref{eq:D1shear}:
\be\label{eq:1shear}
\DD = \frac{\a \om D_0}{\om (1- \a J_0/D_0) + \l} (\Pi_+ + \Pi_-) \ .
\ee
Integrating Eq. (\ref{HL3}) over $\a$ and using $P_1(\s=\pm 1,t) = 0$ leads to:
\be\label{eq:2shear}
\l \Phi = - \frac1{\tel} \Phi - D_0 (\Pi_+ + \Pi_-) - \DD \frac{J_0}{D_0} \ .
\ee
In order to compute $P_1(\s,t)$ simply, one can notice that $H_{\bar\s}(\s,\l;\dot\g) = \int_0^\infty d\tau e^{-\l \tau} G_{\dot\g}(\s,\tau | \bar\s,0)$ obeys the following equation:
\be\nonumber
D_0 H_{\bar\s}''(\s,\l;\dot\g) -\dot\g H_{\bar\s}'(\s, \l; \dot\g) - \l H_{\bar\s}(\s, \l; \dot\g) = - \delta(\s-\bar\s) \ ,
\ee
with boundary condition $H_{\bar\s}(\s = \pm 1, \l; \dot\g )= 0$. We will denote $\k_\pm$ the two real roots of the equation:
\be
D_0 \k^2 -\dot\g \k - \l = 0 \ ,
\ee
and $\Delta = \k_+ - \k_-$.
The solution then reads, respectively for $\s > \bar\s$ and $\s < \bar\s$:
\be
H_{\bar\s}(\s,\l;\dot\g) = A_\pm({\bar\s}) e^{\k_+ \s} + B_\pm({\bar\s}) e^{\k_- \s} \ ,
\ee
with the following equations coming from boundary conditions:
\be\begin{split}
&A_\pm e^{\pm \k_+} = - B_\pm e^{\pm \k_-} \Rightarrow B_\pm = - A_\pm e^{\pm \Delta} \ , \\
&(A_+ - A_-) e^{\k_+ {\bar\s}} + (B_+ - B_-) e^{\k_- {\bar\s}} =0 \ , \\
&\k_+ (A_+ - A_-) e^{\k_+ {\bar\s}} + \k_- (B_+ - B_-) e^{\k_- {\bar\s}} = -\frac1{D_0} \ .
\end{split}\ee
Solving for $A_\pm$ yields:
\be
A_\pm({\bar\s}) = \frac{e^{\mp \Delta - \k_+ {\bar\s}} - e^{- \k_- {\bar\s} }}{2D_0 \Delta \sinh \Delta}.
\ee
The relevant quantities here are $H_{\bar\s}'(\s = \pm 1,\l;\dot\g)$, which are expressed as:
\be\nonumber
\begin{split}
H_{\bar\s}'(\s = \pm 1,\l;\dot\g) &= \k_+ A_\pm({\bar\s}) e^{\pm \k_+} + \k_- B_\pm({\bar\s}) e^{\pm \k_-} \\
&= A_\pm({\bar\s}) \Delta  e^{\pm \k_+} \ .
\end{split}\ee
Using
\be\nonumber
\begin{split}
&D_0 P_0''(\s) = \dot\g P_0'(\s) - J_0 \delta(\s) \ , \\
&P_0'(\s > 0) = - A \zeta e^{-\zeta(1-\s)}, \qquad P_0'(\s < 0) =  A \zeta e^{\zeta \s}   \ ,
\end{split}\ee
in Eq.~(\ref{eq:P1shear}), one finds two contributions. The one coming from $\delta(\bar\s)$ reads:
\be
\Pi_\pm^{\delta} = \pm \left(\frac{J_0 \DD}{D_0}  + \frac1{\tel} \Phi \right) \frac{e^{\pm \k_-} - e^{\pm \k_+}}{2D_0 \sinh \Delta} \ ,
\ee
while the continuous part gives:
\be
\Pi_\pm^{c} = \mp \frac{\dot\g \DD}{D_0} \Delta  e^{\pm \k_+} \int_{-1}^{+1} {\rm d}{\bar\s} P_0'({\bar\s}) A_\pm({\bar\s}) \ .
\ee
From these we deduce:
\be\label{eq:3shear}
\begin{split}
\Pi_+^{\delta} + \Pi_-^{\delta} &= \left(\frac{J_0 \DD}{D_0}  + \frac1{\tel} \Phi \right) \frac{\sinh \k_- - \sinh \k_+}{D_0 \sinh \Delta} \ , \\
\Pi_+^{c} + \Pi_-^{c} &=  -\frac{J_0 \DD \z}{D_0^2 \k_+ \k_-} \frac{(e^{-\k_+}- 1)(e^{-\k_-}- 1)}{e^{-\z} +1} \times \\
&\times\frac{\k_+ \sinh \k_- - \k_- \sinh \k_+}{\sinh \Delta} \ .
\end{split}\ee
We now have all the elements to obtain the solvability condition of the three equations \eqref{eq:1shear}, \eqref{eq:2shear},
and \eqref{eq:3shear}, relating $\DD, (\Pi_+ +\Pi_-)$ and $\Phi$, in terms of 
$\widehat \l = \l \tel$, $\widehat \om = \om \tel$ and $\widehat \g=\dot\g \tel$. 
We obtain the condition
\beq\begin{split}
&\frac{\widehat{\l}}{\widehat\om}\left(1-\frac{\widehat{\omega}-1}{\widehat{\l}+1}\right)
 = 
\frac{-\zeta}{e^{-\zeta}+1}
\frac{(e^{- \k_-}-1)(e^{-\k_+}-1)}{\k_+ \k_-} \times \\& \times\frac{\k_+\sinh{\k_-}-\k_-\sinh{\k_+}}{\sinh{(\k_-)}-\sinh{(\k_+)}} \ .
\end{split}\label{nonzerosol}\eeq

The location of the oscillatory instability in the $(\alpha, \widehat{\gamma})$ plane, as predicted by the above equation, forms a kind of 
``bubble'' region as shown in Fig. ~\ref{fig:1}.
We see that for all $\alpha\in [1/2,\alpha_c]$ where the zero-shear state is unstable, there exists a critical shear-stress $\dot \g_c$ beyond
which the flow is stabilized and becomes laminar. Conversely, coming from high shear, there is a shear-stress $\dot \g_c$ below which the 
flow becomes intermittent, or stick-slip. This stick-slip phenomenon only exists if the elastic coupling between elements is neither too large, nor too
small.

\section{Numerical results}

The linear stability analysis derived in the previous sections is in very
good agreement with numerical simulations of the model with a discrete number of elements.
In order to obtain numerical results we take a set of $N$ elements
characterized by continuous stress variables $\{\sigma_i(t)\}_{i=1\cdots N}$ and boolean variables $\{y_i(t)\}_{i=1\cdots N}$
(indicating if the element is jammed or is fluidized). 
At the beginning of each time step the $\sigma_i$ of jammed elements are first independently updated by generating Gaussian 
distributed increments with mean $-\dot{\gamma} \de t$ and round mean square $\sqrt{2 \de tD_t}$,
where $\de t$ is a (sufficiently small) discretization time step.
All elements for which $|\sigma_i(t)|>\sigma_c$ release their stress with probability $\de t/\tpl$, in which case they
contribute to the $\Gamma_{t+1}$ term in Eq.~\eqref{eq:2} and become fluid.
Finally, fluid elements have a probability $\de t/\tel$ of being re-activated with stress $\sigma_i=0$.
In the limit of large $N$ the probability $P(\sigma,t)$ of finding a particle with stress between $\sigma$ and $\sigma + d\sigma$
evolves according to Eq.~\eqref{eq:FP}.

\begin{figure}
\includegraphics[width=.9\columnwidth]{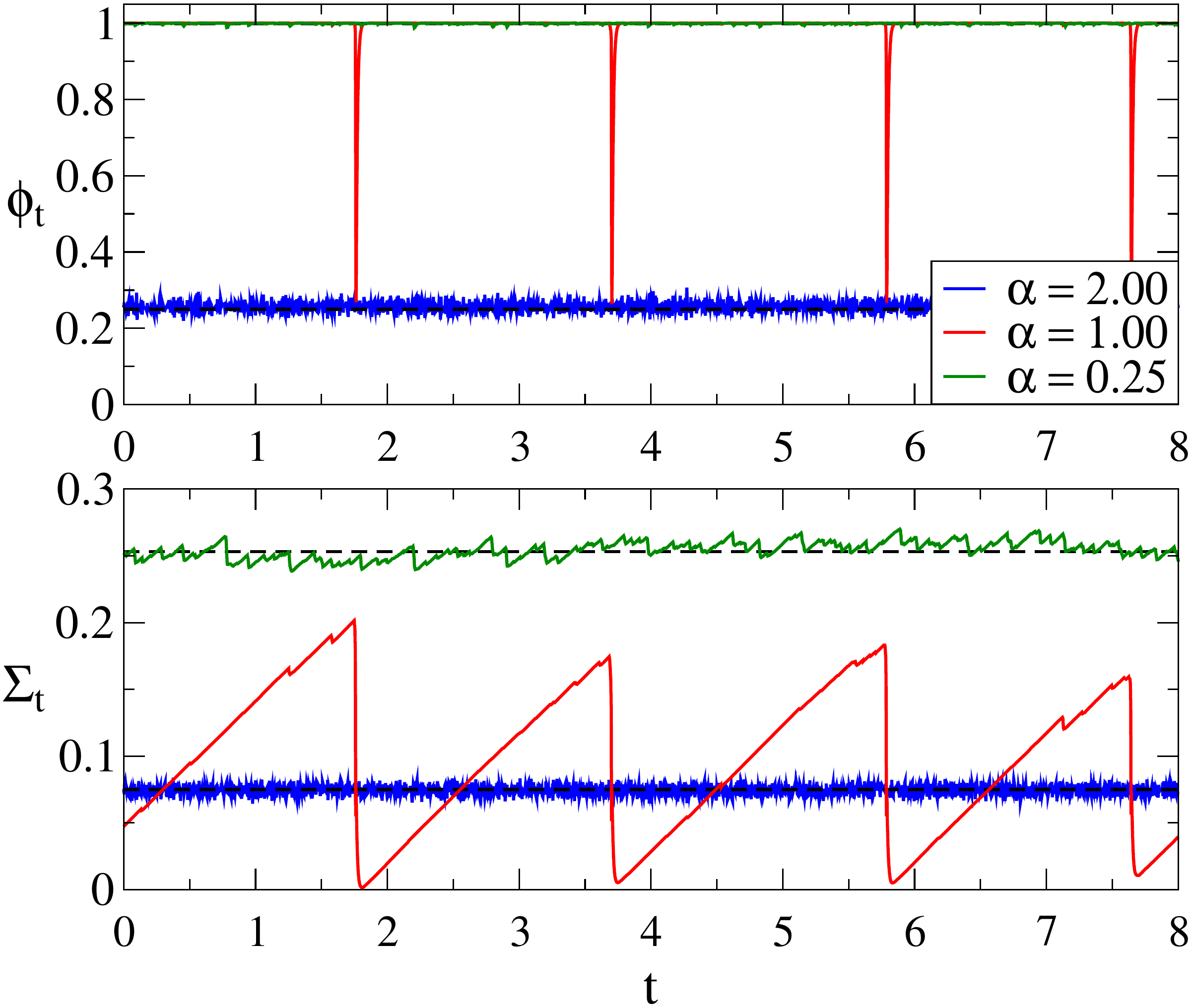}
\caption{
\emph{Top:} Fraction of jammed elements versus time for a population of $N=10^4$ elements undergoing the random walk process described by Eq.~\eqref{eq:FP} (see details in the text).
\emph{Bottom:} corresponding total stress as defined in Eq.~\eqref{eq:stress}.
Different colors correspond to different values of the coupling constant $\alpha$ (black dashed lines
correspond to the analytical stationary solutions in the stable phase).  
Other parameters are: $\wh{\gamma}=0.001$, $\wh{\omega}=15$, $D_{\rm i}=10^{-3}$ , $\de t=10^{-6}$ and $\tpl=0$.}
\label{fig:2}
\end{figure}

In Fig.~\ref{fig:2} we plot the fraction of jammed elements $\phi_t$ versus time 
as well as the total stress $\Sigma_t$ for three different values of the coupling
constant $\alpha$ with $\wh{\omega}=15$, $\tpl=0$, $\wh{\gamma}=0.001$.
One clearly observes that the liquid and jammed phases are separated by a stick-slip regime for intermediate values of the coupling constant $\alpha$, as predicted 
analytically.  A more extensive numerical exploration of the model (not shown here) shows a very good agreement with analytical results for the location of phase 
boundaries depicted in Fig.~\ref{fig:1}. 

\begin{figure}
\includegraphics[width=.9\columnwidth]{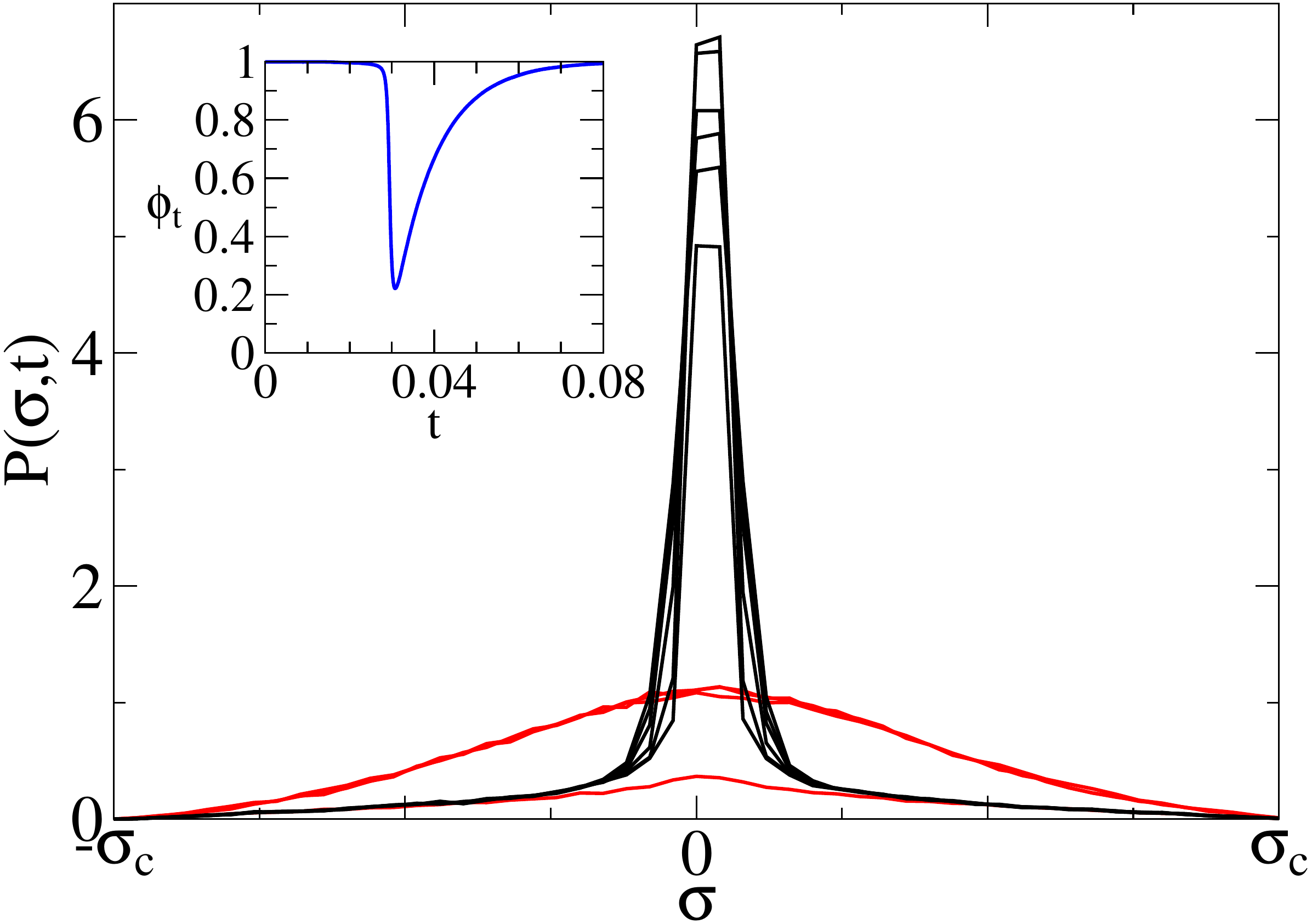}
\caption{Empirical distribution of stress variables $\sigma_i$ for a population of $N=10^5$ elements with zero shear rate. 
Different lines are relative to different snapshots taken at equally spaced time intervals of duration $0.01$ (the corresponding value
for the fraction of jammed elements as a function of time is plotted in the inset).
Red lines correspond to the build up of the instability (the lower line is at $t=0.03$ for example, approximately corresponding to the minimum
of $\phi(t)$ in the inset) while black lines correspond to elements progressively relaxing at $\sigma=0$.
For illustrative purposes we set here $\wh{\gamma}=0$ while other parameters are: $\alpha=1$, $\tel=0.01$, $D_{\rm i}=0.01$, 
$\de t=10^{-6}$, $\wh{\omega}=30$ and $\tpl=0$.}
\label{fig:3}
\end{figure}

In order to have a better understanding of the dynamics of the model in the unstable regime 
it is instructive to look at the empirical probability distribution $P(\sigma,t)$ at different times
corresponding to the accelerated build-up of stress and its subsequent relaxation. 
This is done in Fig.~\ref{fig:3} where we plot the empirical $P(\sigma,t)$ at snapshots equally spaced in time
(see the caption for details). We observe that initially, $P(\sigma,t)$ is peaked around zero. Then, as some elements
become unstable, the stress diffusion constant increases, bringing more elements close to $\sigma_c$. This feedback
can become explosive and lead to a ``spike'' in the number of elements that become unstable simultaneously, see 
Fig.~\ref{fig:3}, inset.

Finally, since all the calculations above where made assuming $\tpl = 0$, we have explored numerically how the instability 
behaves when $\tpl >0$. In this case, we find that the qualitative features of Fig.~\ref{fig:1} are left unchanged as long as 
$\tpl <\tpl^*$ where the value of $\tpl^*$ depends on other parameters, whereas the instability disappears altogether when 
$\tpl > \tpl^*$. For example, in the zero shear rate case, we find for 
$\wh{\omega}=30$ and $\alpha=2$ that the instability disappears at $\tpl^* \approx 0.857\ \tel$. The reason is quite simple: 
since jammed elements become unstable at random times with a dispersion $\sim \tpl$, a large $\tpl$ maims the synchronisation 
mechanism that leads to the oscillating instability discovered here. Since the HL model corresponds
to $\tpl / \tel \to \infty$, this oscillating instability would not have shown up in the original HL setting. 
Note that similarly, a small $\omega$ (corresponding to a large dispersion in the arrival time of the stress signal) is preventing
the instability to occur.

\section{Conclusion} 

In this paper, we have revisited the now classic H\'ebraud-Lequeux model for the rheology of jammed materials. We have argued that a possibly important
time scale is missing from HL's initial specification, namely the characteristic time for a fluidized element to re-jam (chosen to be zero in HL),
whose importance has also been stressed in Refs.~\cite{coussot,MBB12,JLR14}. A linear
stability analysis of the generalized HL model shows that the steady-state solution is in fact unstable for a wide range of parameters, and leads to 
intermittent, stick-slip flows under a constant shear rate. The stick-slip motion disappears when the shear-rate is large (as expected), or when the time for a jammed 
element to become unstable is large compared to the re-jamming time, or else when the elastic coupling between jammed elements 
is either too weak or too strong (see Fig.~\ref{fig:1} above). 

The instability we find is akin to the synchronization transition of coupled elements that arises in many different contexts 
(neurons, fireflies, financial bankruptcies, etc.), see~\cite{St03,GBCTZ15} and references therein. 
Similar instabilities are found  in some variants of the SGR models~\cite{cates1} and within simple phenomenological 
constitutive equations for shear-thickening materials~\cite{cates4}, although the underlying physical mechanism does not seem to be related to the synchronisation
effect discussed here. We hope that our scenario could shed light on the commonly observed intermittent, serrated flows of glassy materials under shear.
It would also be quite interesting to study our instability in finite dimensions (rather than in the HL mean-field limit), and along the lines recently 
suggested by Lin and Wyart~\cite{LW15}, where the HL diffusion term is replaced by a long-range, L\'evy flight diffusion. It is not immediately clear
 whether the stick-slip instability survives such a deep modification of the mathematical and physical nature of the model.

\begin{acknowledgments}
The authors would like to thank J.-L. Barrat, M. E. Cates, A. Rosso, and M. Wyart 
for many stimulating discussions and remarks concerning this work.
\end{acknowledgments}


\bibliographystyle{mioaps}
\bibliography{HL}

\end{document}